\pdfoutput=1

\documentclass[11pt,twoside,a4paper,cmspaper,final,collab]{cms-tdr}
\def\svnVersion{18b5cf1-D}\def\svnDate{2019/10/25}\def\cmsCernNoTag{CERN-EP-2019-294}\def\cmsCernDate{\today}\def\cmsMessage{"Published in Physics Letters B as \href{http://dx.doi.org/10.1016/j.physletb.2019.135049}{\doi{10.1016/j.physletb.2019.135049}.}"}

\begin{document}\cmsNoteHeader{HIN-17-006}
\providecommand {\nch}      {\ensuremath{N_{\mathrm{ch}}}\xspace}
\providecommand {\dnchdy}   {\ensuremath{\rd N_{\mathrm{ch}}/\rd y}\xspace}
\providecommand {\dnchdeta} {\ensuremath{\rd N_{\mathrm{ch}}/\rd \eta}\xspace}

\providecommand {\hydjet}   {\textsc{Hydjet}\xspace}
\providecommand {\ampt}     {\textsc{Ampt}\xspace}
\providecommand {\epos}     {\textsc{Epos}\xspace}
\providecommand {\eposlhc}  {\textsc{Epos LHC}\xspace}
\providecommand {\starlight}{\textsc{STARlight}\xspace}
\providecommand {\dpmjet}   {\textsc{Dpmjet-III}\xspace}
\providecommand {\hijing}   {\textsc{Hijing}\xspace}

\providecommand {\dr}       {\ensuremath{\Delta r}\xspace}
\providecommand {\vz}       {\ensuremath{v_{z}}\xspace}
\providecommand {\npart}    {\ensuremath{N_{\mathrm{part}}}\xspace}
\providecommand {\avenpart} {\ensuremath{\langle N_{\mathrm{part}} \rangle}\xspace}

\hyphenation{had-ron-i-za-tion}
\hyphenation{cal-or-i-me-ter}
\hyphenation{de-vices}
\RCS$HeadURL$
\RCS$Id$
\newlength\cmsFigWidth
\ifthenelse{\boolean{cms@external}}{\setlength\cmsFigWidth{0.85\columnwidth}}{\setlength\cmsFigWidth{0.4\textwidth}}
\ifthenelse{\boolean{cms@external}}{\providecommand{\cmsLeft}{upper\xspace}}{\providecommand{\cmsLeft}{left\xspace}}
\ifthenelse{\boolean{cms@external}}{\providecommand{\cmsRight}{lower\xspace}}{\providecommand{\cmsRight}{right\xspace}}
\ifthenelse{\boolean{cms@external}}
{\providecommand{\supplmat}{the supplemental material [URL will be inserted by publisher]}}
{\providecommand{\supplmat}{Appendix~\ref{app:centrality-npart}}}

\cmsNoteHeader{HIN-17-006}

\title{Pseudorapidity distributions of charged hadrons in xenon-xenon collisions at $\sqrtsNN = 5.44\TeV$}

\date{\today}

\abstract{
Measurements of the pseudorapidity distributions of charged hadrons produced in
xenon-xenon collisions at a nucleon-nucleon centre-of-mass energy of $\sqrtsNN
= 5.44\TeV$ are presented. The measurements are based on data collected by the
CMS experiment at the LHC. The yield of primary charged hadrons produced in
xenon-xenon collisions in the pseudorapidity range $\abs{\eta} < 3.2$ is
determined using the silicon pixel detector in the CMS tracking system. For the
5\% most central collisions, the charged-hadron pseudorapidity density in the
midrapidity region $\abs{\eta} < 0.5$ is found to be $1187 \pm 36\syst$, with a
negligible statistical uncertainty. The rapidity distribution of charged
hadrons is also presented in the range $\abs{y} < 3.2$ and is found to be
independent of rapidity around $y = 0$. Existing Monte-Carlo event generators
are unable to simultaneously describe both results. Comparisons of
charged-hadron multiplicities between xenon-xenon and lead-lead collisions at
similar collision energies show that particle production at midrapidity is
strongly dependent on the collision geometry in addition to the system size and
collision energy.
}

\hypersetup{
pdfauthor={CMS Collaboration},
pdftitle={Pseudorapidity distributions of charged hadrons in xenon-xenon collisions at sqrtsNN = 5.44 TeV},
pdfsubject={CMS},
pdfkeywords={CMS, physics, xenon-xenon, hadrons, multiplicity, spectra}}

\maketitle

\providecommand {\nch}      {\ensuremath{N_{\mathrm{ch}}}\xspace}
\providecommand {\dnchdy}   {\ensuremath{\rd N_{\mathrm{ch}}/\rd y}\xspace}
\providecommand {\dnchdeta} {\ensuremath{\rd N_{\mathrm{ch}}/\rd \eta}\xspace}

\providecommand {\hydjet}   {\textsc{Hydjet}\xspace}
\providecommand {\ampt}     {\textsc{Ampt}\xspace}
\providecommand {\epos}     {\textsc{Epos}\xspace}
\providecommand {\eposlhc}  {\textsc{Epos LHC}\xspace}
\providecommand {\starlight}{\textsc{STARlight}\xspace}
\providecommand {\dpmjet}   {\textsc{Dpmjet-III}\xspace}
\providecommand {\hijing}   {\textsc{Hijing}\xspace}

\providecommand {\dr}       {\ensuremath{\Delta r}\xspace}
\providecommand {\vz}       {\ensuremath{v_{z}}\xspace}
\providecommand {\npart}    {\ensuremath{N_{\mathrm{part}}}\xspace}
\providecommand {\avenpart} {\ensuremath{\langle N_{\mathrm{part}} \rangle}\xspace}

\section{Introduction}

Collisions between ultra-relativistic heavy ions are the only known way of
experimentally studying quantum chromodynamics (QCD) matter at high
temperatures and energy densities. The current understanding is that in such
collisions, a state of matter known as the quark-gluon plasma (QGP) is formed
shortly after the initial impact between the nuclei~\cite{Busza:2018rrf}.

The multiplicity and pseudorapidity distributions of the produced charged
particles are key observables that characterise the initial condition and
subsequent hydrodynamic evolution of the QGP~\cite{Romatschke:2017ejr}. The
dependence of the charged-particle multiplicity on the colliding system,
centre-of-mass energy, and collision geometry can provide information about
nuclear shadowing and gluon saturation effects~\cite{Albacete:2014fwa}, as well
as the relative contributions to particle production from hard scattering and
soft processes~\cite{Kharzeev:2000ph}. These observables also provide input for
models of the particle production process~\cite{dEnterria:2011twh}, from which
information about the formation and properties of the QGP can be extracted.

In October 2017, the CERN LHC collided xenon ($\mathrm{Xe}^{129}$) ions at a
nucleon-nucleon centre-of-mass energy of $\sqrtsNN = 5.44\TeV$, marking the
first time ions other than protons and lead ($\mathrm{Pb}^{208}$) have been
circulated in the LHC. This new collision system provides a unique opportunity
to study the dependence of the charged-particle multiplicity on the size of the
matter produced at LHC energies. Previous measurements of charged-particle
multiplicities in copper-copper (CuCu) and gold-gold (AuAu) collisions at RHIC
have been observed to be sensitive to the collision
geometry~\cite{Alver:2007aa}. The XeXe collision data are thus important for
determining if this feature is also present at higher energies. Comparisons of
the data to predictions of models tuned to describe PbPb collision
data~\cite{Deng:2010xg,Pierog:2013ria,Lokhtin:2005px} can also be used to test
the extent to which these models are able to describe other collision systems.

\sloppy In this Letter, measurements of the pseudorapidity density of primary
charged hadrons, \dnchdeta, in the range $\abs{\eta} < 3.2$ are reported for
XeXe collisions delivered by the LHC. Following earlier analyses in
proton-proton collisions at 0.9--13\TeV~\cite{Khachatryan:2010xs,
Khachatryan:2010us,Khachatryan:2010nk,Chatrchyan:2014qka,Khachatryan:2015jna},
proton-lead collisions at 5.02 and 8.16\TeV~\cite{Sirunyan:2017vpr}, and PbPb
collisions at 2.76\TeV~\cite{Chatrchyan:2011pb}, ``primary'' charged hadrons
are defined as prompt charged hadrons and decay products of all particles with
proper decay length $c\tau < 1$\unit{cm}, where $c$ is the speed of light in
vacuum and $\tau$ is the proper lifetime of the particle. Contributions from
prompt leptons, decay products of longer-lived particles, and secondary
interactions are excluded.

The results are compared to a measurement by the ALICE
Collaboration~\cite{Acharya:2018hhy} and to predictions from the \eposlhc
v3400~\cite{Werner:2005jf,Pierog:2013ria}, \hydjet 1.9~\cite{Lokhtin:2005px},
and \ampt 1.26t5~\cite{Lin:2004en} event generators. The \epos generator is
based on Gribov--Regge theory~\cite{Gribov:1968fc,Drescher:2000ha} and includes
the effect of collective hadronisation in hadron-hadron scattering. The \hydjet
generator treats a heavy ion collision as a superposition of a hydrodynamically
parametrised soft component and a hard component resulting from multi-parton
fragmentation. The \ampt generator combines the \hijing event
generator~\cite{Wang:1991hta} with Zhang's parton cascade
procedure~\cite{Zhang:1997ej} and the ART model~\cite{Li:1995pra} for the last
stage of parton hadronisation.

\section{The CMS detector}

The central feature of the CMS apparatus is a superconducting solenoid of
6\unit{m} internal diameter, providing a magnetic field of 3.8\unit{T}. Within
the solenoid volume are a silicon pixel and strip tracker covering the range
$\abs{\eta} < 2.5$, a lead tungstate crystal electromagnetic calorimeter, and a
brass and scintillator hadron calorimeter, each composed of a barrel and two
endcap sections. Forward calorimeters (HF), made of steel and quartz-fibres and
located on either side of the interaction point, extend the pseudorapidity
coverage provided by the barrel and endcap detectors to $\abs{\eta} < 5.2$.
Muons are detected in gas-ionisation chambers embedded in the steel flux-return
yoke outside the solenoid. The beam pickup timing for experiments (BPTX)
devices are located around the beam pipe at a distance of 175\unit{m} from the
interaction point on either side and provide precise information on the timing
of the incoming beams. A more detailed description of the CMS detector,
together with a definition of the coordinate system used and the relevant
kinematic variables, can be found in Ref.~\cite{Chatrchyan:2008aa}.

Charged hadrons are reconstructed using the silicon pixel detectors installed
during the Phase 1 upgrade~\cite{CMS:2012sda}, which consist of four concentric
cylindrical shells (layers) in the barrel region (BPIX) and three disks on both
sides of the interaction point in the forward region (FPIX). The BPIX and FPIX
consist of a total of 1184 and 672 modules, respectively, and provide excellent
position resolution with their $100{\times}150\mum$ pixels. In this Letter, the
layers of the BPIX are denoted in increasing order of their radial distance
from the beam axis, \ie the layer closest to the beam axis is referred to as
layer 1, the next closest layer is referred to as layer 2, and so on, while the
disks of the FPIX are referred to in increasing order of their longitudinal
distance from the nominal interaction point.

\section{Event selection}
\label{sec:eventselection}

This analysis is based on approximately 1.36 million events. The average
interaction probability per bunch crossing was 1.8\%. Events are selected in
two stages: (i) online, a coincidence of signals from both BPTX devices and at
least one energy deposit above 3\GeV on either side of the HF are required;
(ii) offline, three energy deposits above 3\GeV on each side of the HF and at
least one reconstructed vertex, according to the tracklet-based vertex
reconstruction method described in Ref.~\cite{Chatrchyan:2011pb}, are required.
A study of noncolliding ion bunches shows that the above requirements are
sufficient to reject all backgrounds not originating from interactions between
xenon ions. Consequently, the contribution of background events from beam,
beam-halo, and cosmic ray sources to the observed yields is negligible.

Contamination from electromagnetic (EM) interactions between xenon ions is
studied using simulated events generated by \starlight 2.2~\cite{Klein:2016yzr}
interfaced with \dpmjet 3.0-5~\cite{Roesler:2000he}, and is estimated to be
around 1\%. The event selection efficiency is estimated by fitting the
distribution of the total transverse energy in the HF calorimeter using a
template extracted from simulated \eposlhc events~\cite{Chatrchyan:2011pb}.
Variations in the fit parameters, as well as other observables correlated with
event activity, are used to determine the uncertainty in this method. In
combination with the contamination rate, an overall value of $95 \pm 3\%$ is
quoted for the event selection efficiency.

Nuclei are extensive objects, and their collisions can be characterised by the
centrality, which is related to the impact parameter of the collision. The
centrality can be estimated from the sum of the transverse energy in the HF
calorimeter~\cite{Chatrchyan:2011sx,Chatrchyan:2011pb}. The distribution of the
total transverse energy, after correcting for the event selection efficiency,
is divided into equal partitions and used to classify events into centralities.
The centrality represents a percentile of the total nuclear interaction cross
section~\cite{Chatrchyan:2011pb}; the most central collisions, \ie the
collisions with the smallest impact parameter, are denoted by lower
percentiles. To minimise the amount of EM contamination, which is concentrated
in the 20\% most peripheral events, the analysis is restricted to events with
centrality in the 0--80\% range, where the event selection is fully efficient.

The event centrality is also related to the number of participating nucleons
\npart, which is determined from a Glauber model
calculation~\cite{Miller:2007ri,Loizides:2017ack}. For this calculation, the
nucleon-nucleon inelastic cross section is taken to be $68.4 \pm
0.5\unit{mb}$~\cite{Loizides:2017ack}, while the nuclear radius, skin depth,
and deformation parameter $\beta_{2}$ of the xenon nucleus are set to $5.36 \pm
0.1\unit{fm}$, $0.59 \pm 0.07\unit{fm}$~\cite{Loizides:2014vua}, and $0.18 \pm
0.02$~\cite{Acharya:2018hhy}, respectively. Simulated \eposlhc events are used
to account for the energy resolution of the HF calorimeters and fluctuations in
event activity, which smear the centrality distributions. The resulting values
and associated uncertainties for \npart are listed in \supplmat.

\section{Analysis}

The measurement of \dnchdeta is performed using tracklets, which are pairs of
pixel clusters from two different layers (disks) of the silicon pixel detector.
Pairs of pixel clusters that are produced by the same charged particle have
small differences in $\eta$ and azimuthal angle $\phi$ with respect to the
primary vertex. These correlations are exploited in the analysis to reconstruct
tracklets that reflect the original distribution of primary charged hadrons.
The vertex and tracklet reconstruction algorithms are described in
Ref.~\cite{Chatrchyan:2011pb}.

Six possible types of tracklets can be formed from distinct combinations of the
four layers of the BPIX. In addition, three types of tracklets can be formed
from unique combinations of the three disks of the FPIX. The individual
measurements from all nine combinations are averaged and symmetrised about
$\eta = 0$ to obtain the final results. The different combinations are also
useful for layer-by-layer systematic checks, as they have different
sensitivities to the particle momentum spectrum. Particles with \pt above
40\MeV can be reconstructed using the two BPIX layers closest to the beam pipe.

\begin{figure}[hbt]
  \centering
    \includegraphics[width=0.45\textwidth]{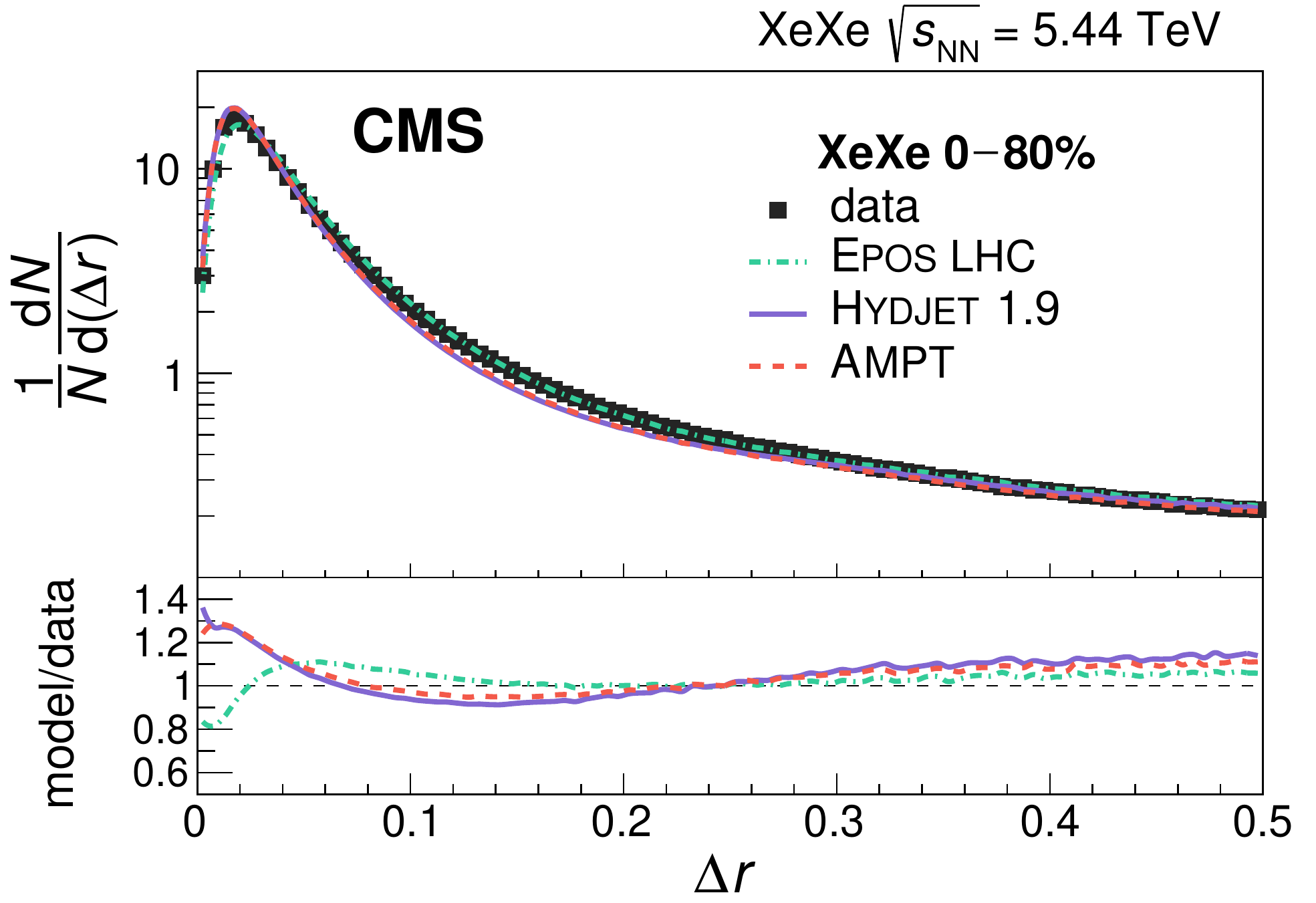}
    \caption{The \dr distributions for tracklets reconstructed from the two
    layers of the BPIX closest to the beam pipe. The distributions are
    normalised by the number of tracklets. The spectrum in collision data
    (black squares) is compared to the spectra obtained from fully simulated
    events generated with \eposlhc v3400~\cite{Werner:2005jf,Pierog:2013ria},
    \hydjet 1.9~\cite{Lokhtin:2005px}, and \ampt 1.26t5~\cite{Lin:2004en}. The
    ratio of the distributions in simulation to data is shown in the bottom
    panel. The statistical uncertainties are smaller than the marker sizes for
    all distributions shown.}
    \label{fig:fig1}
\end{figure}

The angular distance between the two clusters that make up a tracklet is
defined as
\begin{equation}
\dr = \sqrt{{\left(\eta_i - \eta_j\right)^{2} + \left(\phi_i - \phi_j\right)^{2}}},
\end{equation}
where $\eta_{i(j)}$ is the pseudorapidity of the pixel cluster position in the
$i(j)$th layer or disk, calculated with respect to the
primary vertex position, and $\phi_{i(j)}$ is defined similarly for the
azimuthal angle. The \dr distribution for tracklets reconstructed from layers 1
and 2 of the BPIX is shown in Fig.~\ref{fig:fig1}. The spectrum is compared to
fully simulated events generated by \eposlhc, \hydjet, and \ampt.

Tracklets with $\dr < 0.5$ are selected for analysis. This selection criterion
suppresses the combinatorial background from uncorrelated background clusters
and low transverse momentum (\pt) particles that loop around in the high
magnetic field and leave multiple charge deposits per layer of the pixel
detector. The reconstructed tracklet spectrum is then corrected to the
hadron-level event definition by applying a number of correction factors
accounting for the geometric acceptance, reconstruction efficiency, and event
selection efficiency. These correction factors are derived from MC simulations
generated with the aforementioned event generators. The detector response is
simulated with \GEANTfour~\cite{Agostinelli:2002hh} and processed through the
same event reconstruction chain as the collision data. All simulations are
produced with the same vertex distribution along the interaction region as is
observed in data.

Simulations generated with \eposlhc are used as the primary reference for the
derivation of these correction factors because the \dr spectrum obtained from
these simulated events most closely resembles the corresponding spectrum in
data at large \dr, where the combinatorial background is dominant. The other
event generators are used in the study of systematic uncertainties. The
correction factors are calculated as functions of the primary vertex position,
pseudorapidity, and tracklet multiplicity. Typical values of these correction
factors at $\abs{\eta} = 0$ (1.6) range from 1.12 (0.95) at low multiplicities
to 1.01 (0.85) at high multiplicities.

The Jacobian transformation from $\eta$ to rapidity, $y$, can also be derived
from simulations by relating the rapidity density of charged hadrons to the
corresponding pseudorapidity density in each $\eta$
interval~\cite{Abbas:2013bpa,Adam:2016ddh}. The particle composition in data is
assumed to fall within the range of particle compositions predicted by the
various event generators. The final transformation factors applied are the mean
values of the factors derived from each event generator.

The sum of transverse energy in the HF, on which the event selection is based,
is correlated with the charged-hadron multiplicity in the region around $\eta =
0$ where the measurement is made. Hence, the event selection criteria are
susceptible to multiplicity fluctuations and may lead to a nonnegligible bias
in the results~\cite{Morsch:2017brb}. The magnitude of this bias is studied
using various MC event generators by comparing the average \dnchdeta at
midrapidity, defined as $\abs{\eta} < 0.5$, for two sets of generated events:
(i) events selected based on the transverse energy sum in the HF, and (ii)
events selected based on \npart, weighted to have the same distribution of
\npart as the former selection. This provides a comparison of results with and
without the selection bias while also accounting for detector effects that
smear the \npart distribution of selected events. The bias caused by the event
selection criteria is found to be negligible in the centrality interval used in
this analysis.

\section{Systematic uncertainties}

The uncertainties resulting from various systematic effects affecting the
measurement are evaluated. The sources of these systematic uncertainties
include differences between data and simulation for effects such as the
probability of pixel cluster splitting, pixel cluster reconstruction
efficiency, and the fraction of uncorrelated pixel clusters, as well as the
uncertainties in the alignment of pixel detector modules, tracklet selection
criteria, parametrisation of correction factors, consistency between different
tracklet combinations, and model dependence of the correction factors.
Additionally, the uncertainty in the event selection is taken into account as
an independent, fully correlated uncertainty. The individual contributions are
then summed in quadrature to give the total systematic uncertainty.

Pixel cluster splitting refers to when the charge deposit from a single charged
particle is reconstructed as two pixel clusters in close proximity. The
difference in the relative fraction of split clusters between data and
simulation can be estimated by artificially splitting the pixel clusters in
simulation and comparing the resulting modified \dr distribution of cluster
pairs in simulation to that in data. This difference is found to be no more
than 2\%, which results in a variation of 1.8--2.0\% in the \dnchdeta results.
The pixel cluster reconstruction efficiency can be estimated by studying the
fraction of tracklets reconstructed from pixel clusters from the first and
third layers that have a matching pixel cluster in the second layer. The ratio
of this efficiency in data and simulation shows a relative difference of 0.5\%,
which has an effect of 0.5\% when propagated to the final results. The pixel
cluster positions are smeared by the uncertainty in the alignment of the pixel
detector modules, and the effect on the final results is found to be
${<}0.1\%$. The difference in the number of uncorrelated pixel clusters in data
and simulation is estimated by comparing the tracklet \dr distributions in the
region $\dr > 0.3$, where tracklets reconstructed from two uncorrelated
clusters are dominant. Additional pixel clusters (on the order of 1--4\%) were
randomly added to the simulated events such that the tracklet \dr distributions
at large \dr match those in data. A difference of 0.5\% in the final
results is observed at $\eta = 0$, which increases monotonically with
$\abs{\eta}$ to 2.4\% at $\abs{\eta} = 3.2$.

The tracklet selection criteria affect the minimum \pt and signal-to-background
ratio of reconstructed tracklets. The sensitivity of the correction factors to
these effects is checked by varying the nominal selection criterion on \dr by
$\pm 0.1$. The effect of such variations on the final results is found to be
about 0.2\%. The multiplicity variable used in the parametrisation of the
correction factors can be changed to be the number of pixel clusters, which is
independent of the tracklet reconstruction efficiency. The effects of such a
change are negligible. In any given $\eta$ range, measurements can be made
using multiple tracklet combinations. The maximum deviation of the measurements
obtained using each tracklet combination from the final averaged and
symmetrised result, which ranges from 1.0 to 2.1\% within $\abs{\eta} < 1.4$
and up to 5.0\% at larger values of $\abs{\eta}$, is quoted as a systematic
uncertainty. The model dependence of the correction factors is studied by using
different sets of correction factors derived from \hydjet and \ampt, which have
different descriptions of the particle production mechanisms. The predicted
particle spectra and composition can differ significantly among the event
generators, which affect the correction for leptons and the extrapolation of
the measured tracklet spectra to $\pt = 0$. The maximum deviation from the
nominal results is quoted as an uncertainty, and ranges from 2.0--2.2\% within
$\abs{\eta} < 1.0$ to a maximum of 5.0\% around $\abs{\eta} = 2.0$. The model
dependence of the Jacobian transformation from $\eta$ to rapidity is also
evaluated in a similar manner, and the maximum deviation, which ranges from
0.5\% around $\abs{\eta} = 1.4$ to 2.1\% (2.5\%) around $\abs{\eta} = 0$ (3.2),
is quoted as an additional uncertainty for the \dnchdy results.

The determination of event centrality depends on the hadronic event selection
efficiency, as well as the amount of contamination from EM processes. Since the
inefficiency is limited to the most peripheral collisions, the effect of the
uncertainty in the event selection efficiency is to shift the events into other
centrality intervals. Hence, to evaluate the uncertainty in the final results,
different sets of centrality calibrations, derived after varying the event
selection efficiency by its uncertainty, are used to categorise the data. This
leads to a difference of 0.4--25.7\% in the final results, largest in the
75--80\% centrality interval and decreasing towards more central collisions,
which is fully correlated across different centrality intervals and $\eta$
values. The uncertainties in the \npart values are determined by propagating
the uncertainties in the parameters of the Glauber model, which are listed in
Section~\ref{sec:eventselection}, and which range from 0.7 to 8.9\%.

A summary of the systematic uncertainties is given in Table~\ref{tab:tab1}.
With the exception of the uncertainties in the event selection efficiency and
the \npart values, the systematic uncertainties are largely independent of
centrality and highly correlated point-to-point in the region $\abs{\eta} <
1.4$, where only combinations of BPIX layers contribute to the result.

\begin{table}[htbp!]
  \centering
    \topcaption{Sources of systematic uncertainty affecting the measurement of
    charged hadron multiplicities and \avenpart in XeXe collisions at $\sqrtsNN
    = 5.44\TeV$.}
    \label{tab:tab1}
    \begin{tabular}{lc}
    \hline
      Source                                          & [\%]          \\
    \hline
      Pixel cluster splitting                         & 1.8--2.0      \\
      Pixel cluster reconstruction efficiency         & 0.5           \\
      Alignment uncertainty                           & ${<}0.1$      \\
      Uncorrelated pixel clusters                     & 0.5--2.4      \\
      Tracklet selection                              & 0.2           \\
      Tracklet reconstruction efficiency              & ${<}0.05$     \\
      Consistency between tracklet combinations       & 1.0--5.0      \\
      Model dependence                                & 2.0--5.0      \\
      Model dependence (Jacobian transformation)      & 0.5--2.5      \\
      Event selection efficiency (0--5\% to 75--80\%) & 0.4--25.7     \\
      Glauber model calculation                       & 0.7--8.9      \\
    \hline
    \end{tabular}
\end{table}

\section{Results}

\begin{figure}[hbt!]
  \centering
    \includegraphics[width=0.45\textwidth]{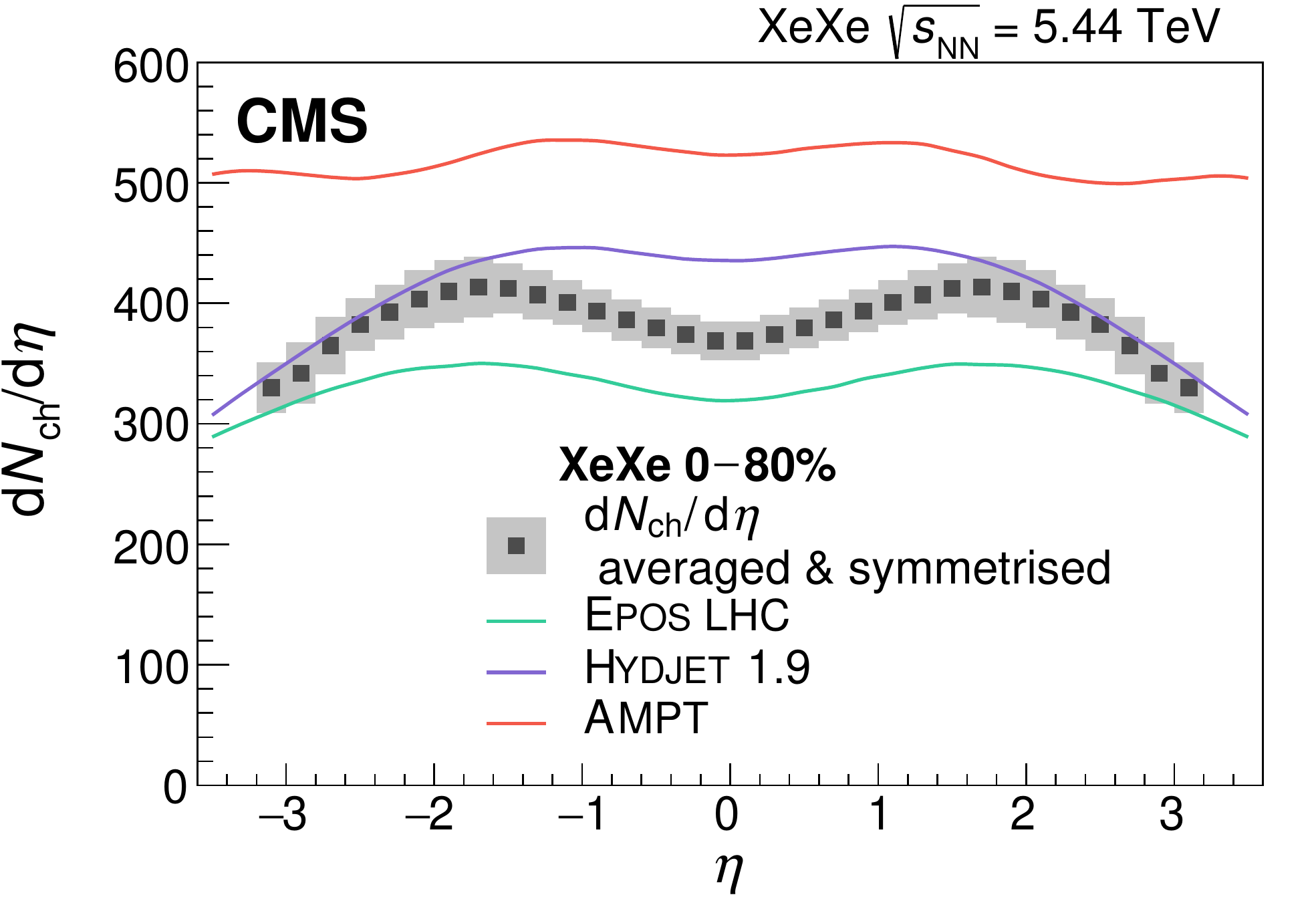}
    \includegraphics[width=0.45\textwidth]{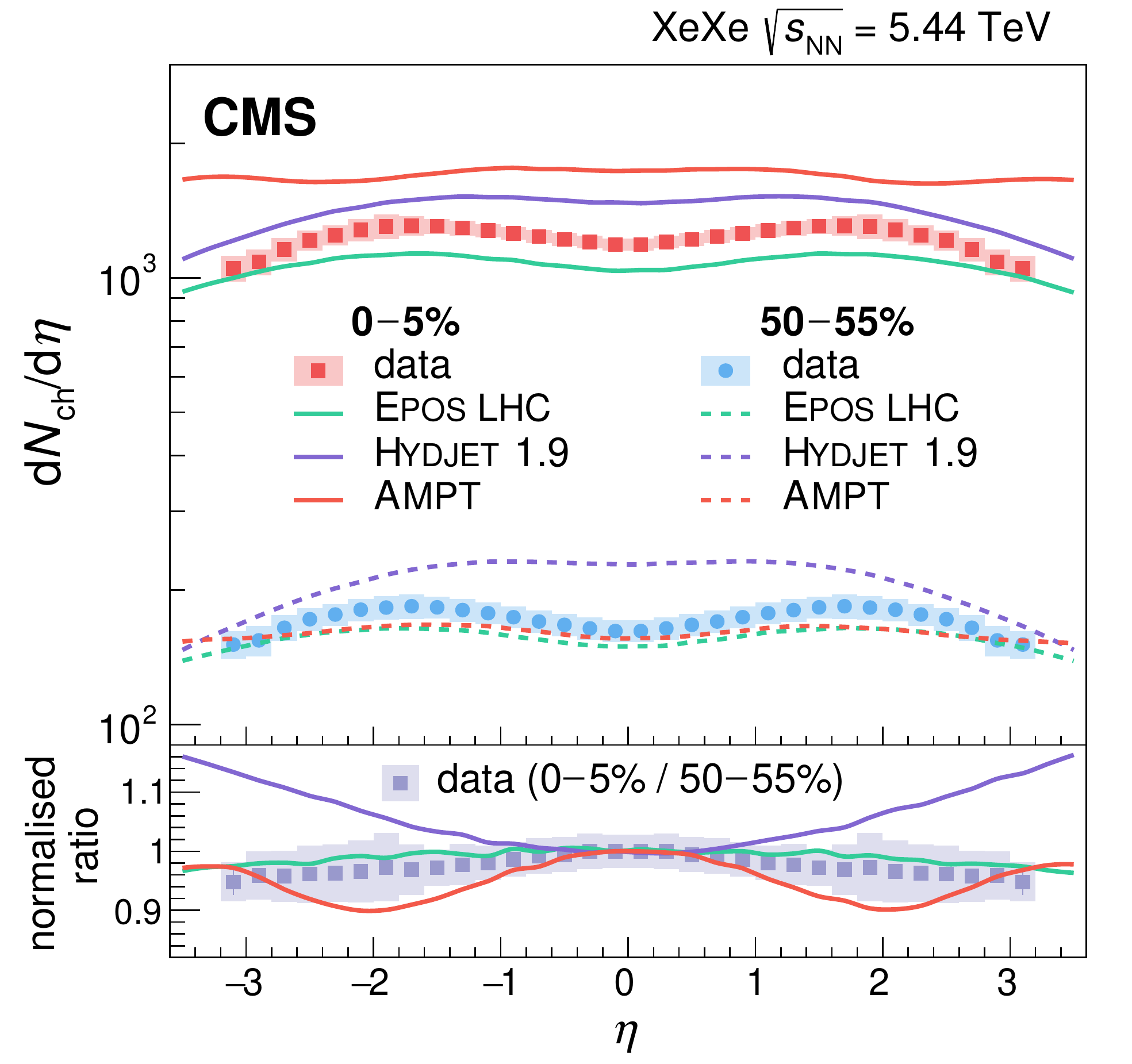}
    \caption{Averaged and symmetrised \dnchdeta distributions in XeXe
    collisions at $\sqrtsNN = 5.44\TeV$ (grey squares), for events in  the
    0--80\% centrality interval (\cmsLeft), as well as the 0--5\% (red squares)
    and 50--55\% (blue circles) centrality intervals (\cmsRight). Predictions
    from the \eposlhc v3400~\cite{Werner:2005jf,Pierog:2013ria}, \hydjet
    1.9~\cite{Lokhtin:2005px}, and \ampt 1.26t5~\cite{Lin:2004en} event
    generators are also shown for comparison. The ratios of the \dnchdeta
    distributions for events in the 0--5\% to those in the 50--55\% centrality
    interval, normalised to unity at midrapidity, are shown in the bottom
    panel. The bands around the data points denote the total systematic
    uncertainties, while the statistical uncertainties are negligible.}
    \label{fig:fig2}
\end{figure}

The pseudorapidity distributions of charged hadrons for $\abs{\eta} < 3.2$ are
shown in Fig.~\ref{fig:fig2} (\cmsLeft) for events in the 0--80\% centrality
interval, and in Fig.~\ref{fig:fig2} (\cmsRight) for events in the 0--5\% and
50--55\% centrality intervals. The bottom panel in Fig.~\ref{fig:fig2}
(\cmsRight) shows the ratios of the \dnchdeta distributions for events in the
0--5\% centrality interval to those in the 50--55\% centrality interval,
normalised to unity at midrapidity. There is a hint of a centrality dependence
in the shape of the \dnchdeta distribution, in that the distribution in
peripheral collisions is flatter than that in central collisions.

None of the event generators are able to fully describe the \dnchdeta
distributions in the three centrality intervals shown, in particular the
\dnchdeta at midrapidity. However, the shapes of the distributions, where the
overall normalisations are factored out, are consistent with those predicted by
the \eposlhc event generator within the total systematic uncertainties. The
centrality dependence of the shape of the \dnchdeta distributions is described
well by \eposlhc but not by the other event generators, as shown in the bottom
panel of Fig.~\ref{fig:fig2} (\cmsRight).

\begin{figure}[hbt!]
  \centering
    \includegraphics[width=0.45\textwidth]{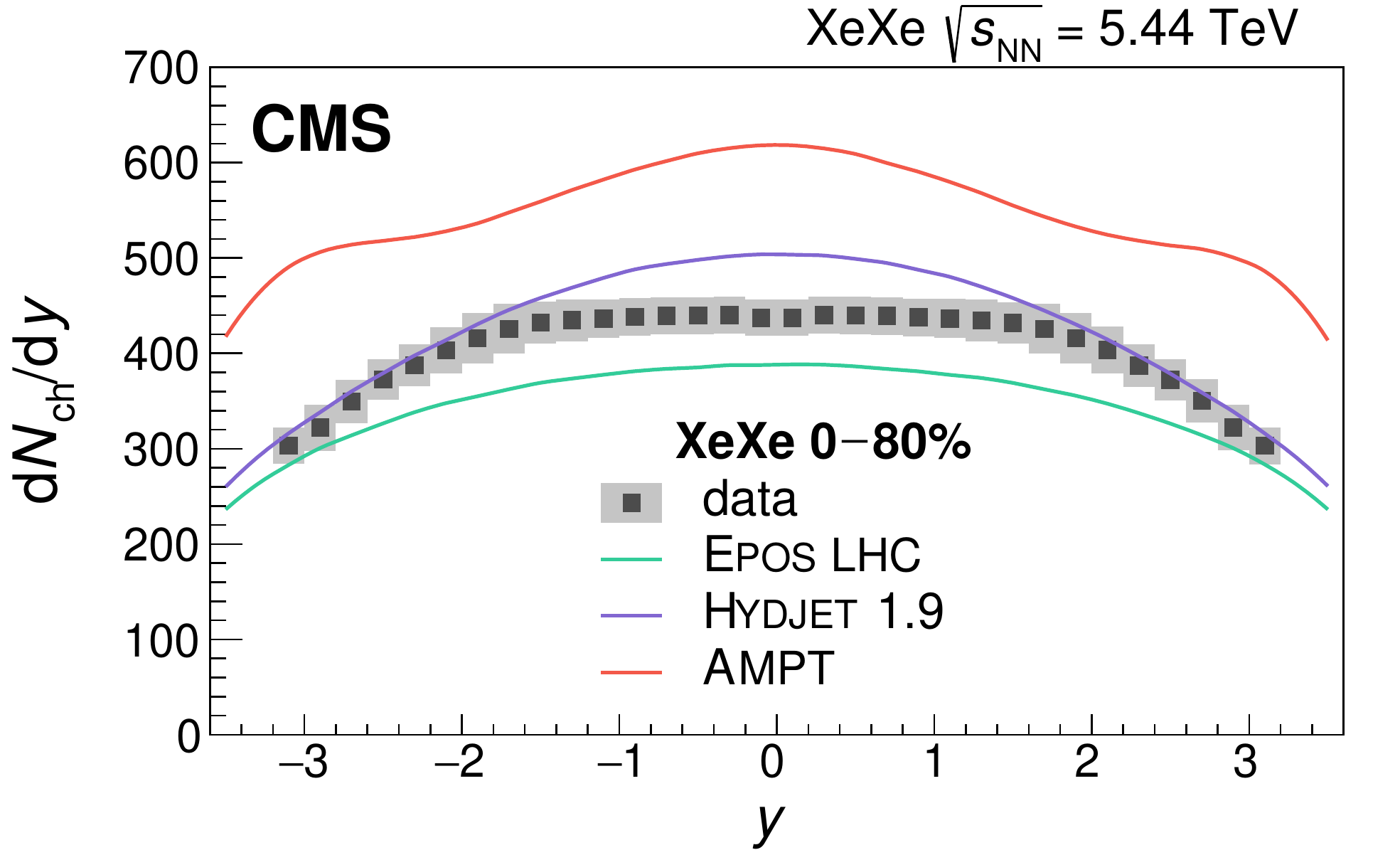}
    \caption{Averaged and symmetrised charged-hadron \dnchdy distribution in
    XeXe collisions at $\sqrtsNN = 5.44\TeV$ for events with 0--80\% centrality
    (grey squares). The band around the data points denotes the total
    systematic uncertainties, while the statistical uncertainties are
    negligible. Predictions from the \eposlhc v3400~\cite{Werner:2005jf,
    Pierog:2013ria}, \hydjet 1.9~\cite{Lokhtin:2005px}, and \ampt
    1.26t5~\cite{Lin:2004en} event generators are also shown in comparison.}
    \label{fig:fig3}
\end{figure}

The rapidity distribution of charged hadrons in XeXe collisions with 0--80\%
centrality is shown in Fig.~\ref{fig:fig3}. The \dnchdy distribution in data is
observed to be consistent with a rapidity plateau in the region $\abs{y} < 1$.
The \dnchdy distributions obtained from the \eposlhc, \hydjet, and \ampt event
generators are also shown for comparison. None of the event generators describe
the plateau around $y = 0$.

\begin{figure}[hbt!]
  \centering
    \includegraphics[width=0.45\textwidth]{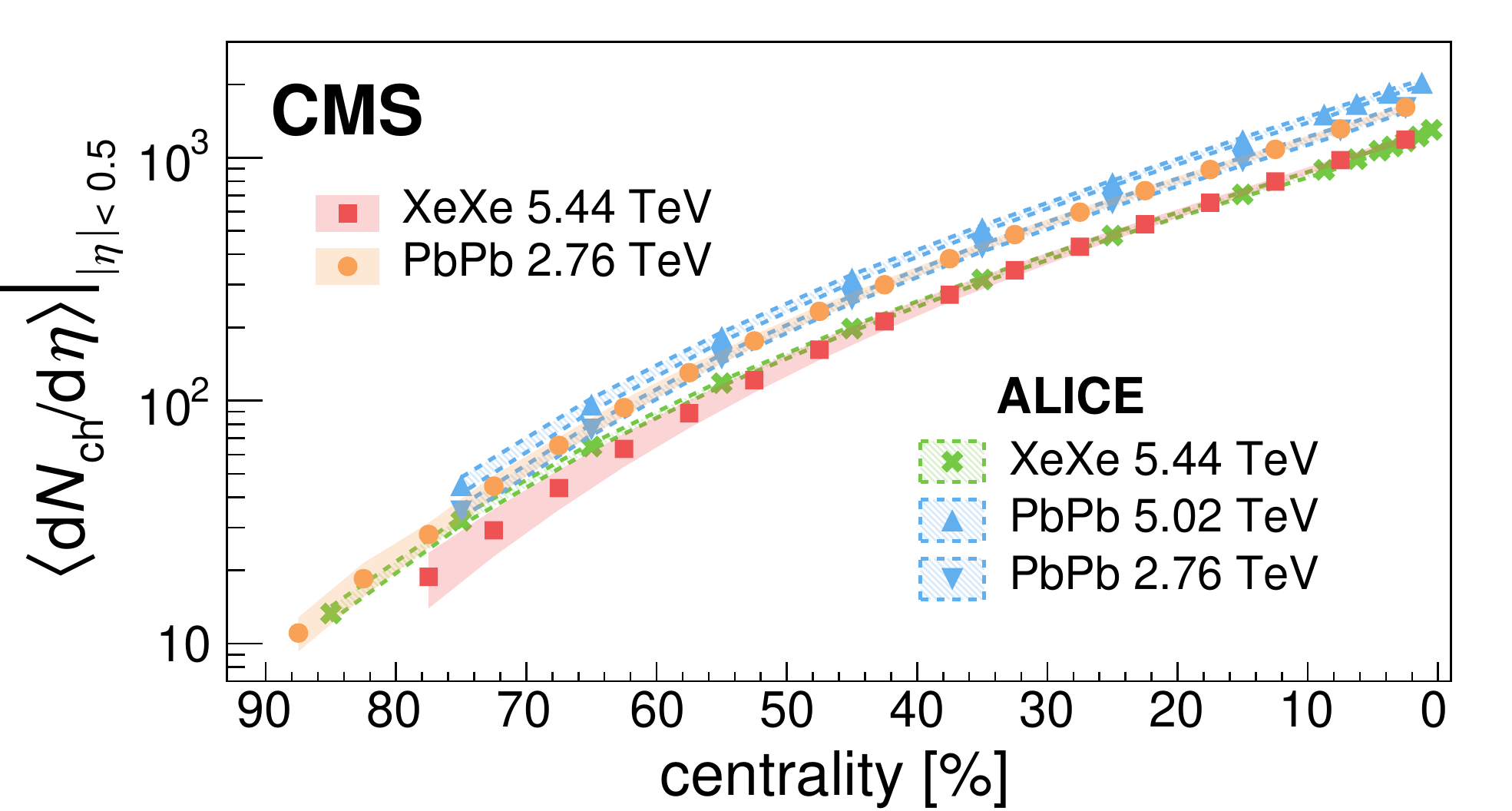}
    \includegraphics[width=0.45\textwidth]{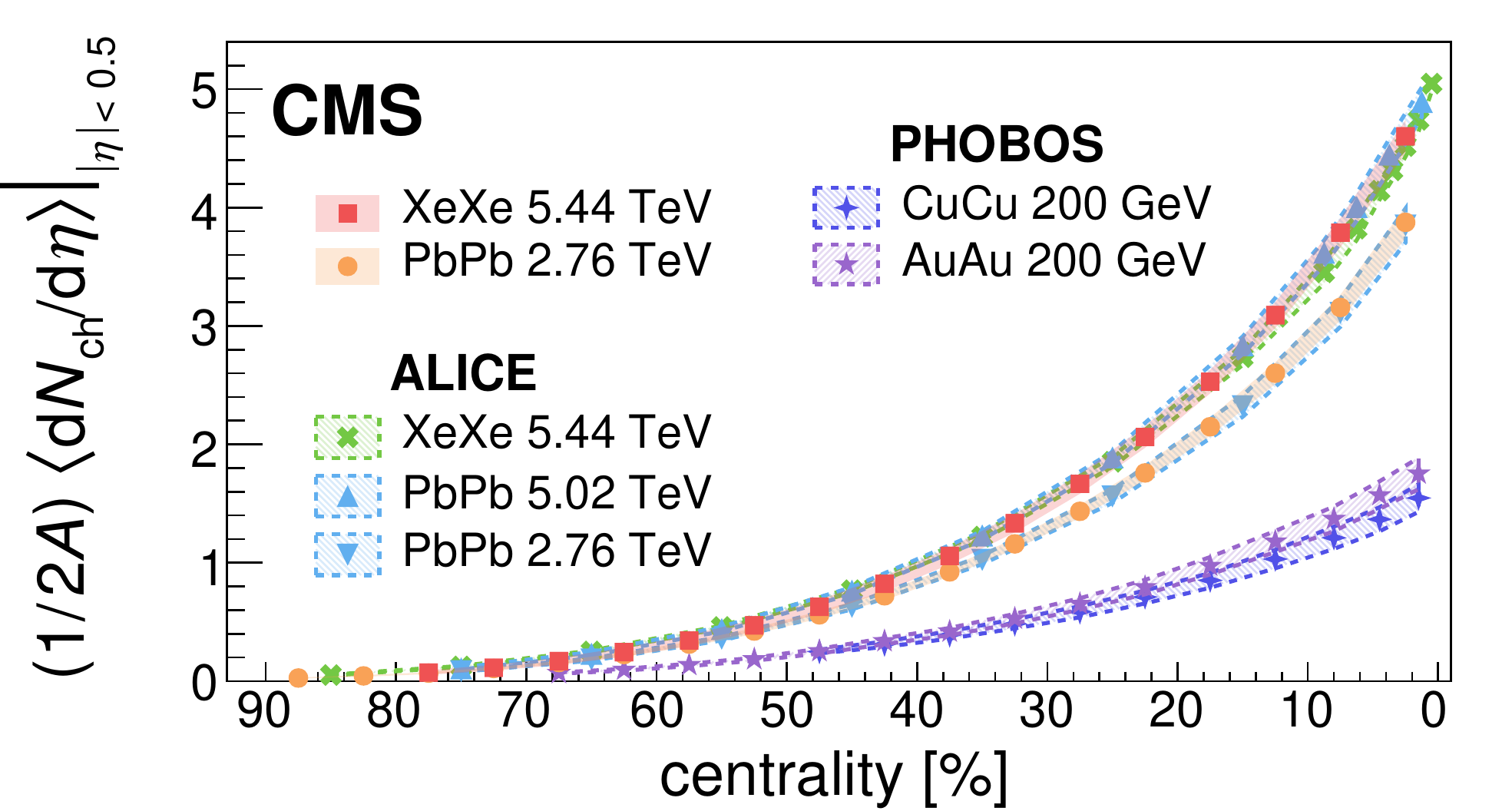}
    \caption{Charged-hadron \dnchdeta in XeXe collisions at $\sqrtsNN =
    5.44\TeV$ at midrapidity as a function of event centrality, shown as is
    (\cmsLeft) and normalised by $2A$ (\cmsRight), where $A$ is the atomic
    number of the nuclei. The results are compared to measurements in PbPb and
    XeXe collisions by the CMS~\cite{Chatrchyan:2011pb} and
    ALICE~\cite{Aamodt:2010cz,Adam:2015ptt,Acharya:2018hhy} Collaborations, and
    to measurements in CuCu and AuAu collisions by the PHOBOS
    Collaboration~\cite{Alver:2010ck}. The bands around the data points denote
    the total systematic uncertainties, while the statistical uncertainties are
    negligible.}
    \label{fig:fig4}
\end{figure}

Figure~\ref{fig:fig4} (\cmsLeft) shows the charged-hadron \dnchdeta at
midrapidity as a function of centrality. For events in the 0--5\% centrality
interval, \dnchdeta is found to be $1187 \pm 36\syst$ at midrapidity. This is
nearly a factor of two greater than the interpolated \dnchdeta in proton-proton
collisions at the same energy~\cite{Khachatryan:2010us} after scaling by $A$,
the atomic number of the nuclei. The results are compared to a measurement at
the same energy for charged particles by the ALICE
Collaboration~\cite{Acharya:2018hhy}, which includes leptons in the analysis.
Within the total uncertainties, the measurements are consistent in the 0--60\%
centrality interval, although the ALICE Collaboration reports a slightly higher
\dnchdeta for more peripheral collisions.

The results are also compared to previous measurements in PbPb collisions at
$\sqrtsNN = 2.76$ and 5.02\TeV by the CMS~\cite{Chatrchyan:2011pb} and
ALICE~\cite{Aamodt:2010cz,Adam:2015ptt} Collaborations. As one would expect,
for the same centrality, \dnchdeta increases with energy and system size. It is
interesting to note that for different colliding nuclei at the same energy,
\dnchdeta is proportional to $2A$. This is evident from Fig.~\ref{fig:fig4}
(\cmsRight), where $(\dnchdeta)/2A$ is shown as functions of centrality for a
variety of colliding nuclei and energies. These results show that the feature
observed at lower energies, that the geometry of the colliding systems plays an
important role in determining the production of particles~\cite{Alver:2007aa},
is also present at the much higher LHC energies.

\begin{figure}[hbt!]
  \centering
    \includegraphics[width=0.45\textwidth]{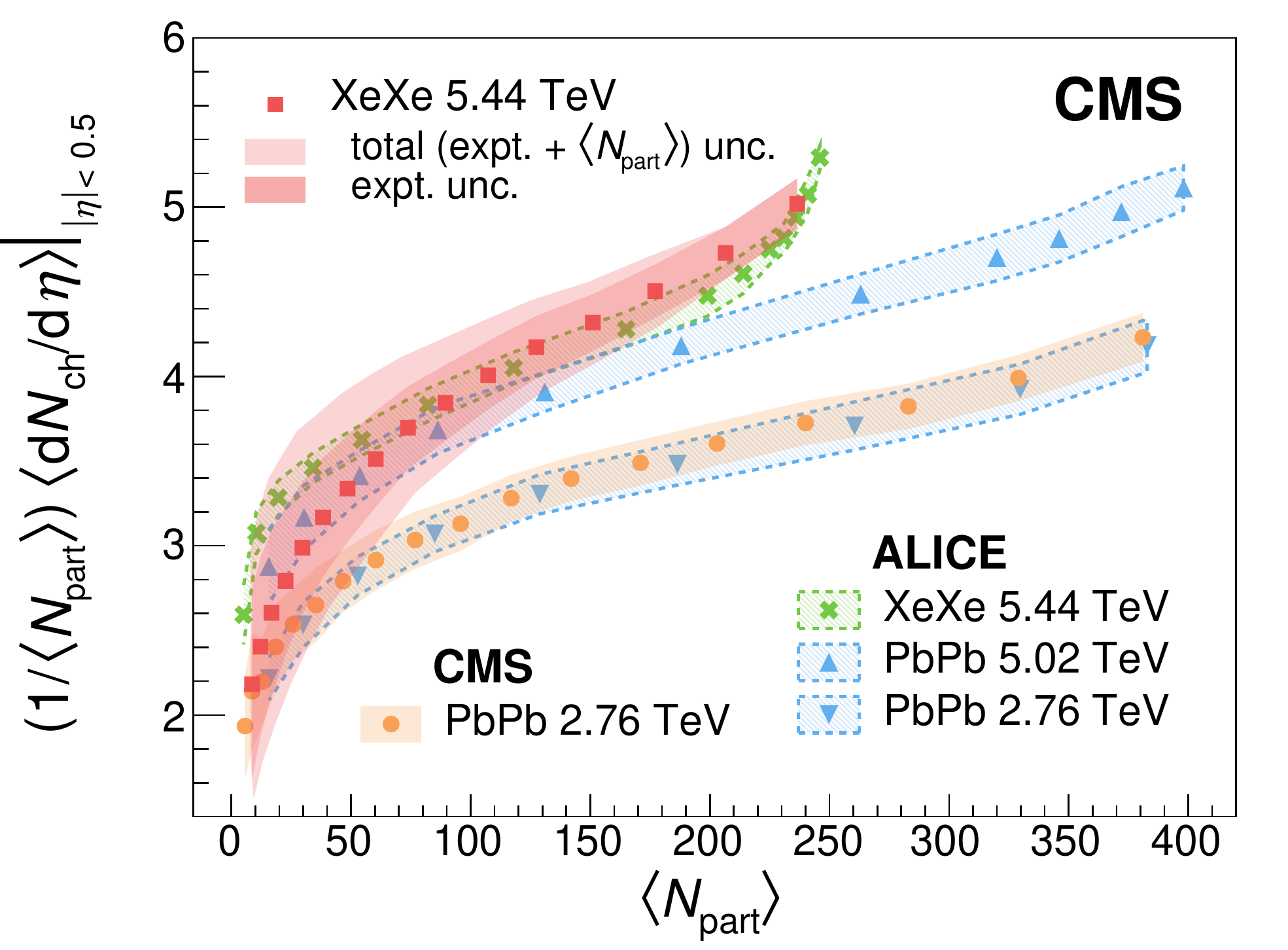}
    \includegraphics[width=0.45\textwidth]{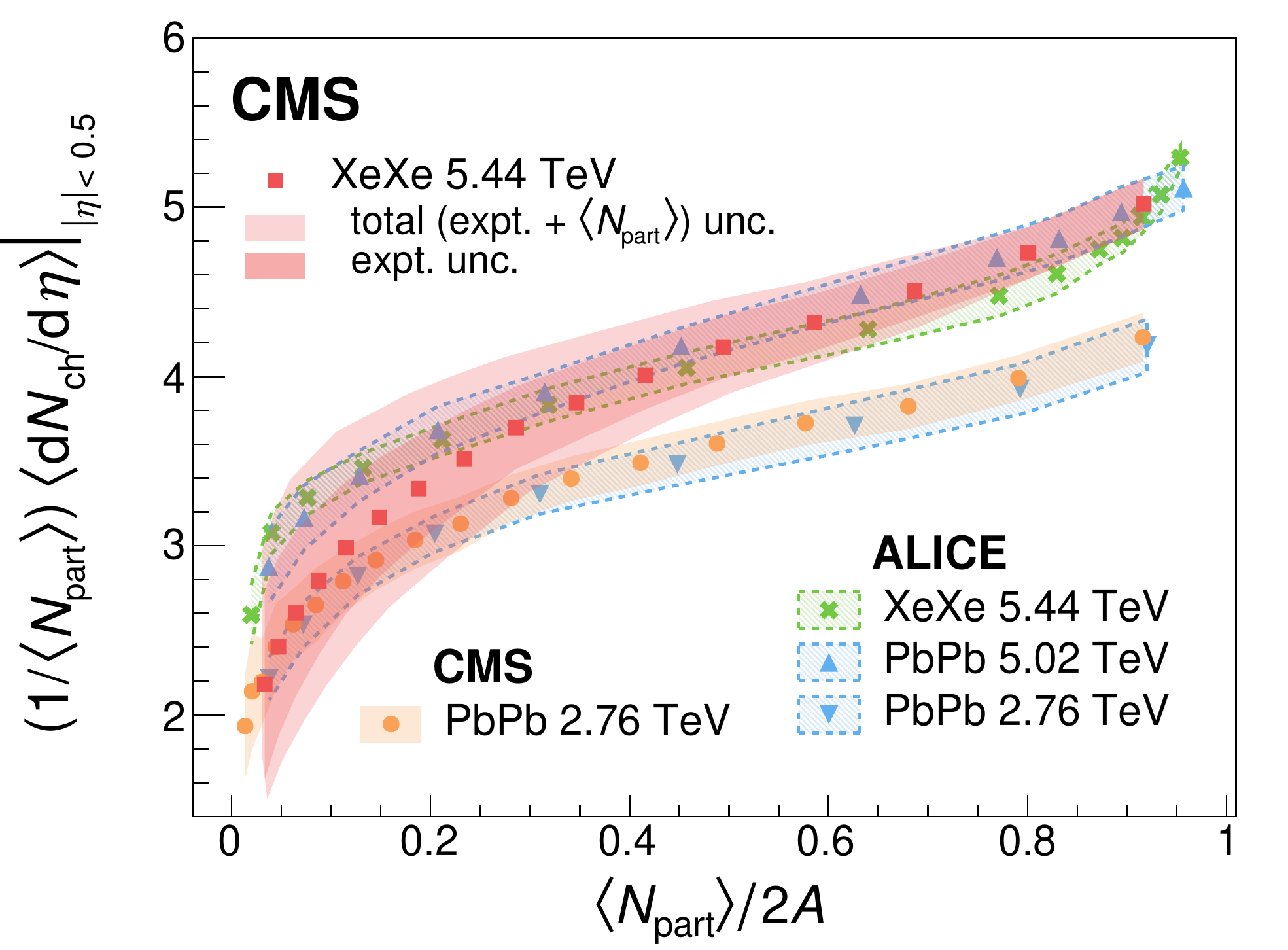}
    \caption{Average \dnchdeta at midrapidity normalised by \avenpart, shown as
    a function of \avenpart (\cmsLeft) and $\avenpart/2A$ (\cmsRight), where
    $A$ is the atomic number of the nuclei. The results are compared to
    measurements in PbPb and XeXe collisions by the CMS~\cite{Chatrchyan:2011pb}
    and ALICE~\cite{Aamodt:2010cz,Adam:2015ptt,Acharya:2018hhy} Collaborations.
    The bands around the data points denote the systematic uncertainties, while
    the statistical uncertainties are negligible.}
    \label{fig:fig5}
\end{figure}

To study the relevance for particle production of the number of participating
nucleons, $(\dnchdeta)/\avenpart$ is shown as a function of \avenpart in
Fig.~\ref{fig:fig5} (\cmsLeft). The results are compared to a measurement at
the same energy by the ALICE Collaboration and to previous measurements in PbPb
collisions at $\sqrtsNN = 2.76$ and 5.02\TeV. As can be seen, the
per-participant multiplicity for XeXe and PbPb collisions with similar
\avenpart, but corresponding to different centrality classes in the two
collision systems, are inconsistent. This is most apparent when nearly
completely overlapping XeXe collisions (0--5\% centrality or $\avenpart \approx
236$) are compared to PbPb collisions with similar \avenpart, for which the
corresponding centrality is approximately 15--20\%. However, as shown in
Fig.~\ref{fig:fig5} (\cmsRight), where $\avenpart/2A$ is used as a proxy for
centrality (the correspondence between centrality and $\avenpart/2A$ is
tabulated in \supplmat), the per-participant charged-hadron multiplicity for
different colliding nuclei are equal within uncertainties when the geometry
(centrality) and energy of the compared systems are the same.

\begin{figure}[hbt!]
  \centering
    \includegraphics[width=0.45\textwidth]{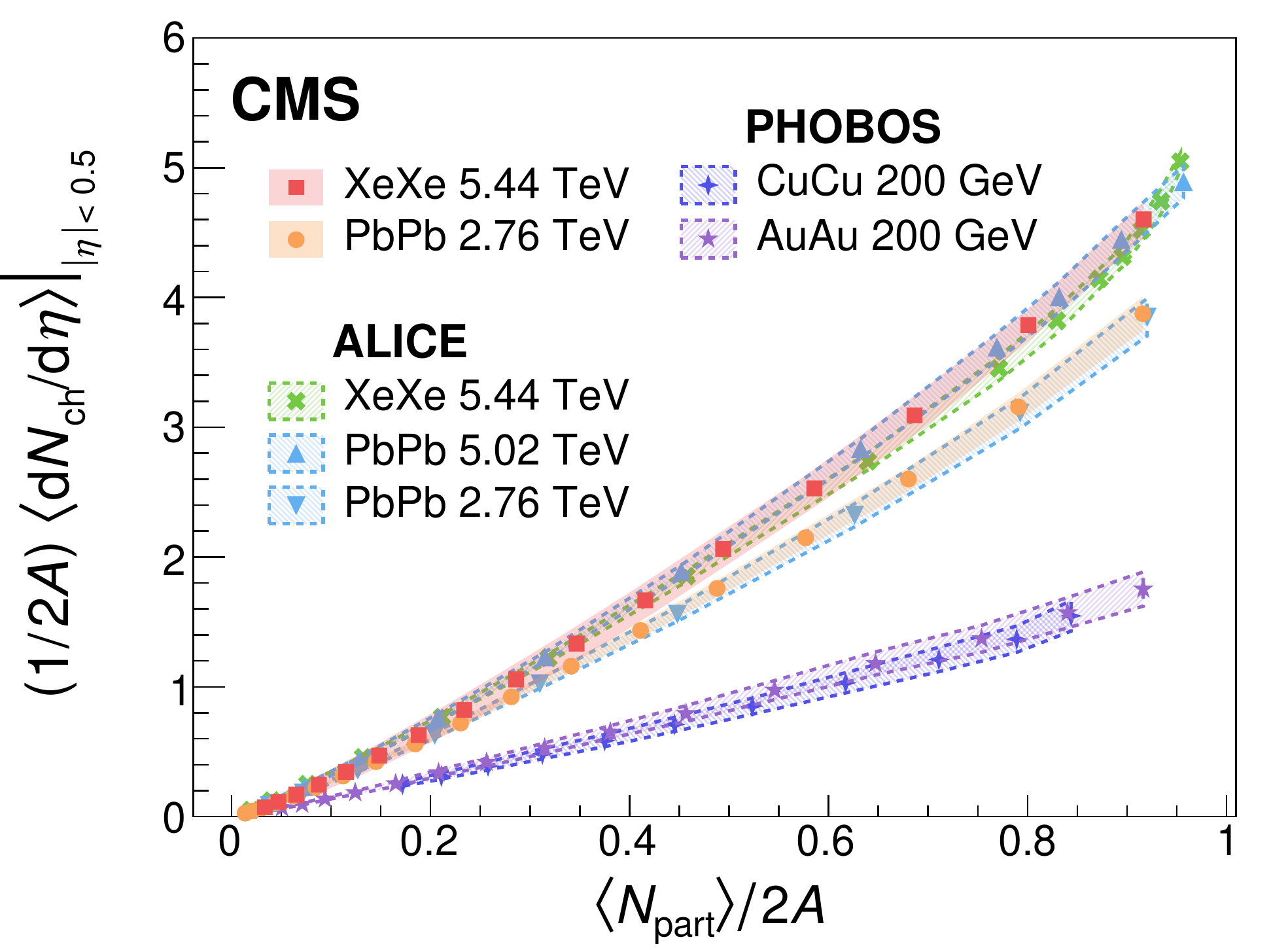}
    \caption{Average \dnchdeta at midrapidity normalised by $2A$, shown as a
    function of $\avenpart/2A$, where $A$ is the atomic number of the nuclei.
    The results are compared to measurements in PbPb and XeXe collisions by the
    CMS~\cite{Chatrchyan:2011pb} and ALICE~\cite{Aamodt:2010cz,Adam:2015ptt,
    Acharya:2018hhy} Collaborations, and to measurements in CuCu and AuAu
    collisions by the PHOBOS Collaboration~\cite{Alver:2010ck}. The bands
    around the data points denote the systematic uncertainties, while the
    statistical uncertainties are negligible.}
    \label{fig:fig6}
\end{figure}

An equivalent representation of Fig.~\ref{fig:fig5} (\cmsRight) is shown in
Fig.~\ref{fig:fig6}, where $(\dnchdeta)/2A$ is shown as a function of
$\avenpart/2A$. In this form, it is clear that multiparticle production scales
as $2A$ times a function of $\avenpart/2A$, indicating a dependence on both the
system size (given by $2A$) and the geometry of the colliding system
(represented by $\avenpart/2A$). Considering that multiparticle production
processes in heavy ion collisions are highly complex---starting with the
initial impact of the two nuclei, through the creation and evolution of a
relativistic fluid, and followed by a hadronisation and scattering phase---it
is not surprising that the result depends on both the colliding system and
energy, in a nontrivial way.

\section{Summary}

The pseudorapidity distributions of charged hadrons in xenon-xenon collisions
at a centre-of-mass energy of 5.44\TeV per nucleon pair are reported. Using
data taken with the upgraded 4-layer silicon pixel detectors, the
charged-hadron pseudorapidity densities, \dnchdeta, are measured to an extended
$\eta$ range of $\abs{\eta} < 3.2$. For events in the 0--5\% centrality
interval, the \dnchdeta at midrapidity is measured to be $1187 \pm 36\syst$,
with a negligible statistical uncertainty. The results are found to be
consistent with the ALICE Collaboration's measurement. The charged-hadron
rapidity density is also presented, and is found to be consistent with a
rapidity plateau in the region $\abs{y} < 1$. The results are compared to
predictions from the \eposlhc v3400, \hydjet 1.9, and \ampt 1.26t5 event
generators. None of the event generators are able to fully describe the
measurements in terms of the magnitude, pseudorapidity dependence, and
centrality dependence of the \dnchdeta distributions, although \eposlhc
describes the shape well. The per-participant \dnchdeta at midrapidity in XeXe
collisions is observed to rise faster with \npart than in PbPb collisions.
However, when comparing events with similar fractional overlap, the
per-participant \dnchdeta is consistent between the two collision systems. The
results also show that the \dnchdeta at midrapidity is a function of the
collision geometry after normalising by $2A$, where $A$, is the atomic number
of the nuclei. This is observed for a variety of collision systems and
energies, both at RHIC and the LHC, demonstrating that final-state
charged-hadron multiplicities are strongly dependent on the collision geometry.
These results provide important constraints on models and generators which
describe multiparticle production in heavy ion collisions at high energies.
They may also help in the characterisation of the initial conditions of the
quark gluon plasma, which is needed for the understanding of its subsequent
hydrodynamic evolution, as well as the properties of this fluid.

\begin{acknowledgments}
We congratulate our colleagues in the CERN accelerator departments for the excellent performance of the LHC and thank the technical and administrative staffs at CERN and at other CMS institutes for their contributions to the success of the CMS effort. In addition, we gratefully acknowledge the computing centers and personnel of the Worldwide LHC Computing Grid for delivering so effectively the computing infrastructure essential to our analyses. Finally, we acknowledge the enduring support for the construction and operation of the LHC and the CMS detector provided by the following funding agencies: BMBWF and FWF (Austria); FNRS and FWO (Belgium); CNPq, CAPES, FAPERJ, FAPERGS, and FAPESP (Brazil); MES (Bulgaria); CERN; CAS, MoST, and NSFC (China); COLCIENCIAS (Colombia); MSES and CSF (Croatia); RPF (Cyprus); SENESCYT (Ecuador); MoER, ERC IUT, and ERDF (Estonia); Academy of Finland, MEC, and HIP (Finland); CEA and CNRS/IN2P3 (France); BMBF, DFG, and HGF (Germany); GSRT (Greece); NKFIA (Hungary); DAE and DST (India); IPM (Iran); SFI (Ireland); INFN (Italy); MSIP and NRF (Republic of Korea); MES (Latvia); LAS (Lithuania); MOE and UM (Malaysia); BUAP, CINVESTAV, CONACYT, LNS, SEP, and UASLP-FAI (Mexico); MOS (Montenegro); MBIE (New Zealand); PAEC (Pakistan); MSHE and NSC (Poland); FCT (Portugal); JINR (Dubna); MON, RosAtom, RAS, RFBR, and NRC KI (Russia); MESTD (Serbia); SEIDI, CPAN, PCTI, and FEDER (Spain); MOSTR (Sri Lanka); Swiss Funding Agencies (Switzerland); MST (Taipei); ThEPCenter, IPST, STAR, and NSTDA (Thailand); TUBITAK and TAEK (Turkey); NASU and SFFR (Ukraine); STFC (United Kingdom); DOE and NSF (USA).

\hyphenation{Rachada-pisek} Individuals have received support from the Marie-Curie program and the European Research Council and Horizon 2020 Grant, contract Nos.\ 675440 and 765710 (European Union); the Leventis Foundation; the A.P.\ Sloan Foundation; the Alexander von Humboldt Foundation; the Belgian Federal Science Policy Office; the Fonds pour la Formation \`a la Recherche dans l'Industrie et dans l'Agriculture (FRIA-Belgium); the Agentschap voor Innovatie door Wetenschap en Technologie (IWT-Belgium); the F.R.S.-FNRS and FWO (Belgium) under the ``Excellence of Science -- EOS" -- be.h project n.\ 30820817; the Beijing Municipal Science \& Technology Commission, No. Z181100004218003; the Ministry of Education, Youth and Sports (MEYS) of the Czech Republic; the Lend\"ulet (``Momentum") Program and the J\'anos Bolyai Research Scholarship of the Hungarian Academy of Sciences, the New National Excellence Program \'UNKP, the NKFIA research grants 123842, 123959, 124845, 124850, 125105, 128713, 128786, and 129058 (Hungary); the Council of Science and Industrial Research, India; the HOMING PLUS program of the Foundation for Polish Science, cofinanced from European Union, Regional Development Fund, the Mobility Plus program of the Ministry of Science and Higher Education, the National Science Center (Poland), contracts Harmonia 2014/14/M/ST2/00428, Opus 2014/13/B/ST2/02543, 2014/15/B/ST2/03998, and 2015/19/B/ST2/02861, Sonata-bis 2012/07/E/ST2/01406; the National Priorities Research Program by Qatar National Research Fund; the Programa Estatal de Fomento de la Investigaci{\'o}n Cient{\'i}fica y T{\'e}cnica de Excelencia Mar\'{\i}a de Maeztu, grant MDM-2015-0509 and the Programa Severo Ochoa del Principado de Asturias; the Thalis and Aristeia programs cofinanced by EU-ESF and the Greek NSRF; the Rachadapisek Sompot Fund for Postdoctoral Fellowship, Chulalongkorn University and the Chulalongkorn Academic into Its 2nd Century Project Advancement Project (Thailand); the Welch Foundation, contract C-1845; and the Weston Havens Foundation (USA).
\end{acknowledgments}

\bibliography{auto_generated}

\ifthenelse{\boolean{cms@external}}{}{
\clearpage
\appendix
\numberwithin{table}{section}
\section{
\texorpdfstring{Supplementary material: Centrality intervals and corresponding \avenpart values}
{Centrality intervals and corresponding mean Npart values}
\label{app:centrality-npart}}

\begin{table}[htbp!]
  \centering
    \topcaption{Centrality intervals and corresponding \avenpart and
    $\avenpart/2A$ values for XeXe collisions}
    \label{tab:tab2}
    \begin{tabular}{ccc}
    \hline 
      Centrality interval [\%] & \avenpart       & $\avenpart/2A$    \\
    \hline
      0--5                     & $236.5 \pm 1.7$ & $0.917 \pm 0.007$ \\
      5--10                    & $206.6 \pm 1.5$ & $0.801 \pm 0.006$ \\
      10--15                   & $177.1 \pm 1.9$ & $0.686 \pm 0.007$ \\
      15--20                   & $151.1 \pm 2.3$ & $0.586 \pm 0.009$ \\
      20--25                   & $127.5 \pm 2.6$ & $0.494 \pm 0.010$ \\
      25--30                   & $107.3 \pm 2.9$ & $0.416 \pm 0.011$ \\
      30--35                   & $89.5 \pm 3.0$  & $0.347 \pm 0.012$ \\
      35--40                   & $73.8 \pm 3.1$  & $0.286 \pm 0.012$ \\
      40--45                   & $60.3 \pm 3.1$  & $0.234 \pm 0.012$ \\
      45--50                   & $48.6 \pm 3.0$  & $0.188 \pm 0.012$ \\
      50--55                   & $38.3 \pm 2.6$  & $0.148 \pm 0.010$ \\
      55--60                   & $29.7 \pm 2.4$  & $0.115 \pm 0.009$ \\
      60--65                   & $22.7 \pm 1.9$  & $0.088 \pm 0.007$ \\
      65--70                   & $16.8 \pm 1.5$  & $0.065 \pm 0.006$ \\
      70--75                   & $12.2 \pm 1.0$  & $0.047 \pm 0.004$ \\
      75--80                   & $8.62 \pm 0.66$ & $0.033 \pm 0.003$ \\
    \hline
    \end{tabular}
\end{table}

}
\cleardoublepage \section{The CMS Collaboration \label{app:collab}}\begin{sloppypar}\hyphenpenalty=5000\widowpenalty=500\clubpenalty=5000\vskip\cmsinstskip
\textbf{Yerevan Physics Institute, Yerevan, Armenia}\\*[0pt]
A.M.~Sirunyan, A.~Tumasyan
\vskip\cmsinstskip
\textbf{Institut f\"{u}r Hochenergiephysik, Wien, Austria}\\*[0pt]
W.~Adam, F.~Ambrogi, E.~Asilar, T.~Bergauer, J.~Brandstetter, M.~Dragicevic, J.~Er\"{o}, A.~Escalante~Del~Valle, M.~Flechl, R.~Fr\"{u}hwirth\cmsAuthorMark{1}, V.M.~Ghete, J.~Hrubec, M.~Jeitler\cmsAuthorMark{1}, N.~Krammer, I.~Kr\"{a}tschmer, D.~Liko, T.~Madlener, I.~Mikulec, N.~Rad, H.~Rohringer, J.~Schieck\cmsAuthorMark{1}, R.~Sch\"{o}fbeck, M.~Spanring, D.~Spitzbart, A.~Taurok, W.~Waltenberger, J.~Wittmann, C.-E.~Wulz\cmsAuthorMark{1}, M.~Zarucki
% ...
\vskip\cmsinstskip
% ...
1:  Also at Vienna University of Technology, Vienna, Austria\\
% ...
\end{sloppypar}
\end{document}